\documentclass{article}
\usepackage{spconf,amsmath,graphicx,amssymb}
\usepackage{booktabs}
\usepackage{array}
\usepackage{paralist}
\usepackage{subfigure}
\usepackage{xcolor}

\title{End-to-end diarization for variable number of speakers with 
local-global networks and discriminative speaker embeddings}

\name{Soumi Maiti$^{\dagger}$\sthanks{The author performed the work while at Google.} Hakan Erdogan$^{\ddagger}$, Kevin Wilson$^{\ddagger}$, Scott Wisdom$^{\ddagger}$, Shinji Watanabe$^{\sharp}$, John R.\ Hershey$^{\ddagger}$}
\address{$^{\dagger}$ The Graduate Center, CUNY \qquad$^{\ddagger}$ Google Research \qquad $^{\sharp}$ Johns Hopkins University}

\begin{document}
\ninept
\maketitle
\begin{abstract}
We present an end-to-end deep network model that performs meeting diarization from single-channel audio recordings.  End-to-end diarization models have the advantage of handling speaker overlap and enabling straightforward handling of discriminative training, unlike traditional clustering-based diarization methods.   The proposed system is designed to handle meetings with unknown numbers of speakers, using variable-number permutation-invariant cross-entropy based loss functions.  We introduce several components that appear to help with diarization performance, including a local convolutional network followed by a global self-attention module, multi-task transfer learning using a speaker identification component, and a sequential approach where the model is refined with a second stage.  These are trained and validated on simulated meeting data based on LibriSpeech and LibriTTS datasets; final evaluations are done using LibriCSS, which consists of simulated meetings recorded using real acoustics via loudspeaker playback.  The proposed model performs better than previously proposed end-to-end diarization models on these data.  
\end{abstract}
\section{Introduction}
\label{sec:intro}
Diarization is the task of predicting ``who spoke when'' given a recording of, e.g. a meeting or conversation \cite{tranter2006overview,anguera2012speaker,ryant2019second},
and is an important additional step for many speech applications like automatic speech recognition \cite{el2019joint,Chen2020libricss,medennikov2020target}. In this work, we focus on diarization for meeting audio, where there may be overlapping speech, from an unknown but bounded number of speakers. In addition, we focus on continuous speech diarization, where the aim is to diarize recordings without prior information about utterance boundaries. Handling overlapping speech is important: in the ICSI meeting corpus, for example, it has been observed that there are overlaps in speech up to 13\% of the time~\cite{ccetin2006overlap}.  This number may be larger for less formal meetings such as dinner conversations.

Traditionally, speaker diarization uses speaker embeddings and clustering. Speaker clustering diarization~\cite{shum2013unsupervised, sell2014speaker, senoussaoui2013study} is done in multiple steps: segment the audio, extract speaker embeddings, and perform clustering. Usually an i-vector \cite{dehak2010front}, d-vector \cite{wang2018speaker} or x-vector \cite{snyder2019xvector} is used as the speaker embedding. Such speaker verification embeddings assume one speaker at a time is active; hence such a model cannot handle speaker overlap.

In contrast, we follow an emerging trend in using an end-to-end neural network for diarization~\cite{fujita2019, fujita2019end} in a meeting scenario, where diarization is predicted at each frame.  Such a network can be trained to predict diarization from meeting mixtures directly, and hence does not rely on external speaker embeddings.  End-to-end diarization can therefore be trained from audio with speaker overlap. 

We propose a two-stage end-to-end diarization model. First stage uses a time-dilated convolutional neural network (TDCN) to extract local features from speech and a self-attention neural network to focus on global speaker modeling. We also show that a learned speaker identification module further improves diarization performance. 

We show significant improvement over previously proposed end-to-end diarization models on two simulated datasets, with 100K and dynamic mixing of meetings. Additionally, we find that using learned joint speaker embeddings with corresponding losses further improves overall diarization performance.

\textbf{Contributions:}
Our proposed approach makes the following contributions:
\begin{inparaenum}[(1)]
\item an end-to-end  system that handles a variable number of speakers in a single step,
\item an effective architecture consisting of local TDCN followed by a global self-attention network, 
\item a speaker classification auxiliary task to improve the global discriminability of embeddings,
\item the use of linear-complexity self attention to improve performance, and
\item a sequential architecture, that uses a second stage to refine the initial estimates.  
\end{inparaenum}

\section{Related Work}
\label{sec:related}

Traditionally, clustering-based methods are used for speaker diarization, using representations such as i-vectors~\cite{shum2013unsupervised, sell2014speaker}, x-vectors~\cite{garcia2017speaker}, or d-vectors~\cite{wang2018speaker, wan2018generalized}.
Such systems first detect small speech segments, then extract a speaker embedding for each segment, and finally cluster the embeddings. Such clustering based diarization methods are effective only when one speaker is present in each segment, but cannot handle overlapping speech.  
In recent years, some hybrid methods combining clustering and discriminative methods addressed overlap \cite{medennikov2020target, huang2020speaker} but they do not perform end-to-end diarization directly. TS-VAD \cite{medennikov2020target} which uses speaker embeddings as conditioning inputs to a neural network was inspired from a speaker conditioned VAD approach \cite{Ding2020}.

In contrast, end-to-end diarization has been a recent trend motivated by the promise of discriminative training.  Previous works ~\cite{von2019all,  fujita2019, fujita2019end, fujita2020neural,horiguchi2020end, e2ediarizationpatent} have introduced a variety of architectures including BLSTM \cite{fujita2019}, self-attention \cite{fujita2019end} and their combination \cite{fujita2020end} that handle overlapping speech for a fixed numbers of speakers. 

Whereas some approaches handle variable numbers of sources but not overlapping speech, \cite{zhang2019fully,li2019discriminative}, some recent end-to-end approaches \cite{fujita2020neural, von2019all, horiguchi2020end} handle both cases. 
The latter  utilize recursive decoder-style models to estimate each speaker's activity, one at a time
until all the speakers have been decoded.  Theoretically recursive models could handle an unbounded number of speakers, but the recursion may introduce difficulty in training as well as run-time inefficiency.  Our approach avoids these problems by directly diarizing all speakers at once, up to a known maximum number of speakers.

\section{Methods}
\label{sec:technical}
Given an input audio signal $x$, a speaker diarization model estimates a binary speaker activity image 
$\mathbf{y}  \in \{0,1\}^{S \times T}$, with elements $y_{s,t} = 1$ if speaker $s$ is present at segment $t$, and   $y_{s,t} = 0$ otherwise.  $S$ is the maximum number of speakers.  Note that in the case of overlap, $\sum_s y_{s,t} > 1$, and if a speaker $s$ is not present in a meeting, then $y_{s,t} = 0: \forall t$.
We generate diarization labels at a fixed rate of 10 acoustic frames per second. 
End-to-end diarization aims to predict this binary image through diarization probabilities $\mathbf{\hat y} \in [0,1]^{S\times T}$ within a permutation of speakers.

\subsection{Proposed method: end-to-end diarization network}

\begin{figure}[htb]
  \centering
 \includegraphics[width=.7\linewidth]{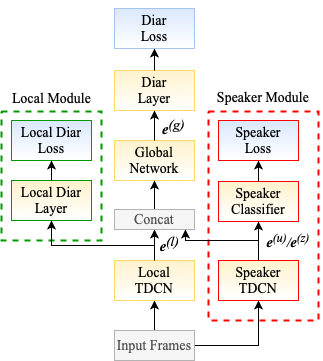}
 \caption{Proposed local-global diarization model, with optional local  and speaker losses; the global network can be LSTM or SA.}
\label{fig:res}
\end{figure}

The proposed diarization model consists of two subnetworks: TDCN (local) and SA (global).
Inspired by recent advances using residual convolutional networks in separation~\cite{Luo2019tasnet}, we use a stacked TDCN as the first diarization submodule with $M$ dilation layers and $4$ repeats with a total of $4M$ blocks.
The local submodule maps input audio $x$ into local embeddings $\mathbf{e}^{(l)}_t\in \mathbb{R}^D$ for each frame.

Since convolutional networks have a limited receptive field, we use a self-attention module to model global context. Self-attention (SA) blocks used are similar to transformer encoder layers~\cite{vaswani2017attention}. Each self-attention block consists of multi-head attention layer with $H$ heads. The output is then fed to two linear layers, where the first layer expands the model dimension ($D$) by a factor of $4$ with a ReLU activation, and the second layer projects back to the model dimension. Residual connections are used around this block, and layer normalization is applied to the output. Multiple such self-attention blocks are stacked. The global layers map local features $\mathbf{e}^{(l)}_t$ to global features $\mathbf{e}^{(g)}_t \in \mathbb{R}^D$. A linear output layer with a sigmoid activation,
called diarization layer in Figure \ref{fig:res},
produces all diarization probabilities $\hat y_{s,t} \in [0,1]$
which are thresholded and post-processed to obtain predicted binary diarization labels.
For typical meetings, the sequence length $T$ can be prohibitively high for self-attention models because their memory and computation complexity have a quadratic dependence on $T$.
The quadratic complexity of the self-attention module arises from the softmax attention computation. For linear approximation, we can replace the softmax attention with a linear dot product of kernel feature map ($\phi$) of query $Q$ and key $K$. 
Given query $Q$, key $K$ and value $V$, where $Q, K, V \in \mathbb{R}^{T\times D}$ the output of the self-attention module can be computed as
\begin{equation}
    O = \operatorname{softmax}\left(\frac{QK^T}{\sqrt{D}}\right) V.
\end{equation}
Inspired by~\cite{katharopoulos2020transformers}, to avoid such complexity we use a linear approximation of full-attention:
\begin{equation}
    O = \phi(Q)(\phi(K)^T V).
\end{equation}
We use the feature map proposed in~\cite{katharopoulos2020transformers}, $\phi(x) =\operatorname{elu}(x)+1$, where the exponential linear unit $\operatorname{elu}(x) = x$ if $x > 0$, and $\alpha(e^x-1)$ if $x \leq 0$ ~\cite{clevert2016elu}.
Such approximation allows us to use a large self-attention model with simpler memory requirements.

\subsection{Permutation Invariant Loss}
\label{ssec:perm_loss}
To allow our end-to-end diarization network to handle permutations between its predictions and the reference labels,
as shown in Figure \ref{fig:res}, we use a diarization loss which is a permutation-invariant cross entropy loss:
\begin{eqnarray}
    L_{\text{diar}}(\mathbf{\hat y}, \mathbf{y}) &=& \min_{\pi \in \Pi} \sum_{s=1}^S \sum_t \text{BCE} (\hat y_{\pi(s),t}, y_{s,t}),
    \label{eq:diar}
    \\
    \text{BCE}(\hat y, y) &=& -\left[y \log(\hat y) + (1-y)\log(1-\hat y)\right],
\end{eqnarray}
where $\pi(s)$ is the permuted index, and  $\Pi$ is the set of permutations of $S$ items. 
For some models, we use a {\em local diarization loss} on top of the embeddings $\mathbf{e}^{(l)}$ obtained from an initial convolutional network. This local loss has the same form as the final diarization loss (\ref{eq:diar}), where the only difference is that it uses local embeddings $\mathbf{e}^{(l)}$ as input, which are processed by a separate diarization layer to obtain a locally estimated diarization probability image.

\subsection{Speaker module}
\label{ssec:speaker}
Though speaker identification is not required for the task of diarization, the neural network has to discriminate between speakers in the meeting.  The network at training time sees a limited amount of meeting audio and so the embeddings may only be trained to contrast with speakers in the local region. Adding a speaker identification loss may encourage embeddings to be globally discriminable. To this end, we propose to train an auxiliary speaker module jointly with the diarization network. The speaker loss minimizes frame-wise speaker identification from a global set of speakers in the training set.
We propose two alternative versions of the speaker module, one producing a joint speaker embedding and another producing multiple individual speaker embeddings per frame. 

In the joint speaker embedding, for each frame we identify active speakers in that frame. The speaker module predicts a speaker label vector, $\mathbf{u}_t \in \mathbb{R}^{C}$, where $C$ is the number of speakers in the training set, and    $ u_{i,t} = 1$  if speaker $i$ is active in frame $t$ and $ u_{i,t} = 0$ otherwise. 
Note that, speaker overlap is indicated by multiple $1$'s in the joint speaker label vector at a frame. 
The speaker network probability output $\hat u_{i,t}$ is generated by passing speaker embedding $\mathbf{e}^{(u)}_t \in \mathbb{R}^D$ through a speaker classifier layer as shown in Figure \ref{fig:res}. The speaker classifier consists of a linear layer with a sigmoid activation, and is trained with binary cross-entropy as:
\begin{equation}
    L_{\text{jointspk}} = \sum_t \sum_i \text{BCE} (\hat u_{i,t}, u_{i,t}).
    \label{eq:jointspk}
\end{equation}

For individual speaker embeddings, we assign a label to each individual speaker at each frame. 
The speaker label is $\mathbf{z}_t \in [C+1]^{S}$, where $S$ is the maximum possible number of speakers in a meeting and $[C+1]$ indicates the set of all integers from $0$ to $C$. We assign each speaker in a meeting a speaker slot from $S$ such slots. At frame $t$, for each speaker $i$ with allotted slot $s$, if speaker $i$ is speaking at frame $t$,  $i$ is assigned and if the speaker is not speaking at $t$, a dummy speaker id (zero) is assigned.
Each label vector is formed as $z_{s,t} = i$ if output slot $s$ is active and assigned to  $i$  at frame $t$ and $z_{s,t} = 0$ otherwise.
This model outputs $S$ small embeddings $\mathbf{e}^{(z)}_{s,t} \in R^{\frac{D}{S}}$ for each frame $t$ from which we derive output probabilities $\hat z_{i,s,t} \in [0,1]$ through $S$ different linear mappings and softmax operations (speaker classifier layer), where $i$ is a training speaker index.  $l_2$ normalization is applied on $\mathbf{e}^{(z)}_{s,t}$.  We let the model decide on the order of speakers predicted and train with a permutation-invariant softmax cross entropy speaker identification loss:
\begin{equation}
     L_{\text{indspk}} = \min_{\pi \in \Pi} \sum_{s,t,i} \delta(z_{s,t}, i) \log(\hat z_{i,\pi(s),t}).
     \label{eq:indspk}
\end{equation}

The speaker module takes the same input as the diarization network. We use a TDCN network to generate embedding ${\bf e}^{(z)} \text{ or } {\bf e}^{(u)}$, 
concatenate with local network embedding $\mathbf{e}^{(l)}$, and feed the result to the self-attention module. For individual speaker embeddings, we concatenate $S$ embeddings $\mathbf{e}^{(z)}_s$ into $\mathbf{e}^{(z)}$ before $l_2$-normalization. For individual speaker embeddings, we make sure the model generates same speaker permutation for diarization and speaker losses. We also experiment with directly applying the joint speaker loss on local embedding ${\bf e}^{(l)}$ instead of extracting speaker embeddings.

\subsection{Sequential model}
We also experimented with a sequential diarization model, where we feed the outputs from a first diarization model into a second one along with the input signal. Diarization probabilities from a first round of output are concatenated with the features calculated from the input signal and used in the second network. This enables the second network to focus on parts where the first network does not get right. We use all the losses in the first and second networks. This sequential application of networks is shown to improve the results.

\section{Experiments}
\label{sec:exp}
\subsection{Simulated meeting data}
\label{ssec:simulated_meeting}
We train end-to-end diarization models with simulated meeting-style audio mixtures. We prepare two separate training and test sets with varying overlap. 
LibriMeet-100K uses utterances from LibriSpeech dataset to create 100,000 120 second meetings with maximum 8 speakers and an overlap ratio between 20\% and 50\%. LibriMeet-Dyn uses utterances from LibriTTS and forms on-the-fly meetings of length 90 seconds involving maximum 8 speakers and with overlaps between 0\% and 40\%. LibriMeet-Dyn has infinite training data and uses on-the-fly mixing which makes it more powerful. Meeting dynamics to generate the meetings are borrowed from LibriCSS \cite{Chen2020libricss} where we draw a random overlap target for each meeting but there are constraints such as no overlapping utterances from the same speaker, and at a given time instant, there can at most be two active speakers.

Input features are 64-dimensional log-mel spectrograms with 40 ms window and 10 ms hop. We concatenate 21 neighboring feature vectors and downsample the result result to a rate of 10 Hz (hop size of 100 ms), which are the input features to the model.

\subsection{Simulated meeting experiments}
We train two baseline end-to-end diarization models, BLSTM and SA. BLSTM was configured with 2 layers and 512 units in each layer. For SA, we use $H=8, D=512$ and $6$ such layers. We train a local TDCN model, with $32$ layers and $M=8$. Our local-global model, has TDCN followed by SA with same architecture as baseline. We also train a TDCN-BLSTM model, where BLSTM is used as global model to compare effectiveness of SA with BLSTM. 
All models are trained with Adam optimizer with learning rate of $10.0^{-4}$ and with batch size 3. For sequential model, we use smaller network TDCN and SA, TDCN-small has $24$ layers and $M=6, D=256$, SA-small has $4$ layers and $H=8$. We also use a large SA model, SA-Large with $H=8, D=512$ and $10$ layers. We train our models with meetings where each one is about $90-120$ seconds long. All models were trained with same input and same computation time. All models except BLSTM were trained for 800K iterations, since BLSTM training speed is slower it was trained for $200K$. We evaluate the systems with diarization error rate (DER) \cite{bredin2019pyannoteaudio}.

\begin{table}[t]
\centering
\scriptsize
\caption{Simulated meeting experiments.}
\begin{tabular}{p{3cm}cc}
\toprule
& \multicolumn{2}{c}{DER(\%)} \\
Model &       LibriMeet-100K & LibriMeet-Dyn \\
\midrule
BLSTM   &   39.3 & 40.0\\
TDCN-BLSTM & 27.8 & 22.1\\ 
SA & 24.9 & 29.8\\
TDCN & 22.1 & 12.5 \\
TDCN-SA & 21.4 & 11.5 \\
TDCN-SA + local loss & 16.3 & 9.4\\
\bottomrule
\end{tabular}
\label{tbl:baseline}
\end{table}

Results are reported in Table \ref{tbl:baseline}. For LibriMeet-100K trained models, we observe SA has lower DER than BLSTM. This is a similar finding as previous papers~\cite{fujita2019end}. We also observe that TDCN only model achieves lower DER than SA. Moreover TDCN-SA model achieves lower DER than TDCN, especially adding diarization loss on both TDCN and SA embeddings, we observe lower DER. We also observe that BLSTM as a global model with TDCN is worse than TDCN. This is probably because the the TDCN-BLSTM model training speed was slower due to BLSTM and the TDCN model may have been under-trained due to slower speed. 

When training with LibriMeet-Dyn, we observe that dynamically mixed meetings achieves lower DER. We observe a similar pattern to LibriMeet-100K training, where TDCN-SA with local loss performs best and achieves a DER of 9. Note that the test sets here are not the same, each training set comes with its own test set, so the DER numbers are not directly comparable between two columns, but the model performances are similar across two training/test setups.

\subsection{Experiments with speaker loss}
Next we add a separate speaker loss module
and train it jointly with the diarization network for the best performing model, TDCN-SA+local, from Table \ref{tbl:baseline}. 
We test using the individual speaker loss, and joint speaker loss on eval sets of LibriMeet-100K and LibriMeet-Dyn. 
Results are in Table \ref{tbl:spk_loss}. Using the speaker loss directly on local embeddings
(second row of Table \ref{tbl:spk_loss}) slightly improves DER for LibriMeet-Dyn, but worsens DER for LibriMeet-100K. Individual speaker loss improves DER by $2.2\%$, and joint speaker loss improves DER by $2.5\%$. DER on LibriMeet-Dyn improves by about $1.3\%$ for both loss types.

\begin{table}[t]
\centering
\scriptsize
\caption{Adding speaker losses to TDCN-SA model + local loss.}
\begin{tabular}{rrcc}
\toprule
Speaker & & \multicolumn{2}{c}{DER} \\
Embedding & Loss &  LibriMeet-100K & LibriMeet-Dyn \\
\midrule
-- & -- & 16.3 & 9.4 \\
Local (${\bf e}^{(l)}$) & $L_\mathrm{jointspk}$ (\ref{eq:jointspk}) & 18.1 & 9.3 \\
Joint (${\bf e}^{(u)}$) & $L_\mathrm{jointspk}$ (\ref{eq:jointspk}) & 13.8 & 8.1 \\
Indiv. (${\bf e}^{(z)}$) & $L_\mathrm{indspk}$ (\ref{eq:indspk}) & 14.1 & 8.1 \\
\bottomrule
\end{tabular}
\label{tbl:spk_loss}
\vspace{-15pt}
\end{table}


\subsection{Linear attention and sequential model experiments}
Since we train using longer audio clips, (few minutes of audio), we often observe limitation in using a large self-attention model. Using linear approximation of self-attention allows us to use a larger SA model with similar memory requirements of a smaller full-attention SA model. 
We report results of linear vs full attention using SA and TDCN-SA models in Table \ref{tbl:linear}.
We observe, SA-Full and SA-Linear achieves similar DER given enough training time, for LibriMeet-100K about $~25\%$ and LibriMeet-Dyn around $~30\%$. SA-Large-linear attention achieves lower DER than SA. When using TDCN and SA together, we observe TDCN-SA-Large-Linear achieves lowest DER in both datasets, about $6\%$ in LibriMeet-Dyn and $12\%$ in LibriMeet-100K. This denotes that using larger self-attention model helps in diarization and for training a long context diarization model, linear approximation of self-attention is useful. The decomposition of DER in terms of missed speech, false acceptance and speaker confusion is as follows. The TDCN-SA-Large+local+speaker model yields 1.6, 1.8, and 8.3 when trained on LibriMeet-100K, and 1.3, 1.7, and 3.2 when trained on LibriMeet-Dyn.

\begin{table}[t]
\centering
\scriptsize
\caption{Full vs linear attention. TDCN-SA use local + speaker loss.}
\begin{tabular}{lccc}
\toprule
 &  &  \multicolumn{2}{c}{DER} \\
Model & Attention &    LibriMeet-100K & LibriMeet-Dyn \\
\midrule
SA & Full & 25.3 & 30.0 \\
SA & Linear & 25.4 & 29.6 \\
SA-Large & Linear & 23.8 & 27.6 \\
TDCN-SA & Full & 13.8 & 8.1 \\
TDCN-SA-Large& Linear & 11.7 & 6.2 \\
\bottomrule
\end{tabular}
\label{tbl:linear}
\end{table}

Table \ref{tbl:sequential_diar} shows results with a sequential model versus a single step model. The sequential model improves DER by $\approx1\%$ absolute.
\begin{table}[t]
\centering
\scriptsize
\caption{Comparison of sequential model vs single-step model.}
\begin{tabular}{p{3cm}lc}
\toprule
Model & Dataset & DER \\
\midrule
TDCN-SA & LibriMeet-100K & 11.7 \\
TDCN-SA(Sequential) & LibriMeet-100K & 11.0\\
TDCN-SA & LibriMeet-Dyn & 6.2\\
TDCN-SA(Sequential) & LibriMeet-Dyn & 5.3 \\
\bottomrule
\end{tabular}
\label{tbl:sequential_diar}
\end{table}

\vspace{-.5cm}
\subsection{Experiments on LibriCSS}
LibriCSS is a dataset of meeting-like data recorded in a conference room with far-field microphones where speech utterances are played from loudspeakers placed in the room \cite{Chen2020libricss}. The order of utterances and overlap amounts are decided using a meeting simulation tool similar to the one we used for our training data. The meetings in this test set are 10 minutes long
and always contain 8 speakers. 
We also have access to the clean sources that were played from the loudspeakers, so in addition to the LibriCSS re-recorded data, we also consider anechoic and artificially reverberated versions, and evaluate the performance of our diarization models on them. The results are presented in Table \ref{tbl:libricss}.

We report the results with the best performing model, a sequential two step model with TDCN-SA-Large and linear attention trained with LibriMeet-Dyn data.
We used a threshold of 0.7 and a median filter of length 31 frames to post-process the output diarization probabilities to obtain the diarization labels. This setup was better than using a threshold of 0.5 and no median filter.

\begin{table}[t]
\centering
\caption{DER on various forms of LibriCSS test data.}
\vspace{2.5pt}
\resizebox{\linewidth}{!}{%
\begin{tabular}{llccccccc}
\toprule
Test data & Training data & 0L & 0S & 10 & 20 & 30 & 40 & avg \\
\midrule
anechoic & LibriMeet-Dyn &  16.4 & 12.5 & 15.8 & 17.1 & 16.8 & 20.0  &  16.4  \\
reverberated & LibriMeet-Dyn & 8.4 & 11.9 & 10.3 & 12.4 & 11.5 & 10.8  &  10.9 \\
re-recorded & LibriMeet-Dyn & 12.4 & 15.9 & 9.6 & 11.8 & 17.1 & 14.0  &  13.5 \\
\bottomrule
\end{tabular}
}
\label{tbl:libricss}
\end{table}
The best result is obtained on artificially reverberated version which shows that the diarization model relies considerably on the reverb structure to determine the diarization output. Since we use single channel data, the information coming from the reverb is the consistent RIR filter used for a speaker over the duration of the meeting which seems important to get better results. We conjecture that due to this reliance as well as due to mismatch with the training data which is always reverberated, the dry/anechoic mixture gets the worst performance among the three conditions.
The performance was not significantly varying with overlap ratio as long as the training data included examples covering the range of overlap ratios appearing in the test set.

\vspace{-.17cm}
\subsection{Variable number of speakers}
\begin{figure}[htb]
\centering
\includegraphics[width=4.8cm, trim={5  5 5 5},clip]{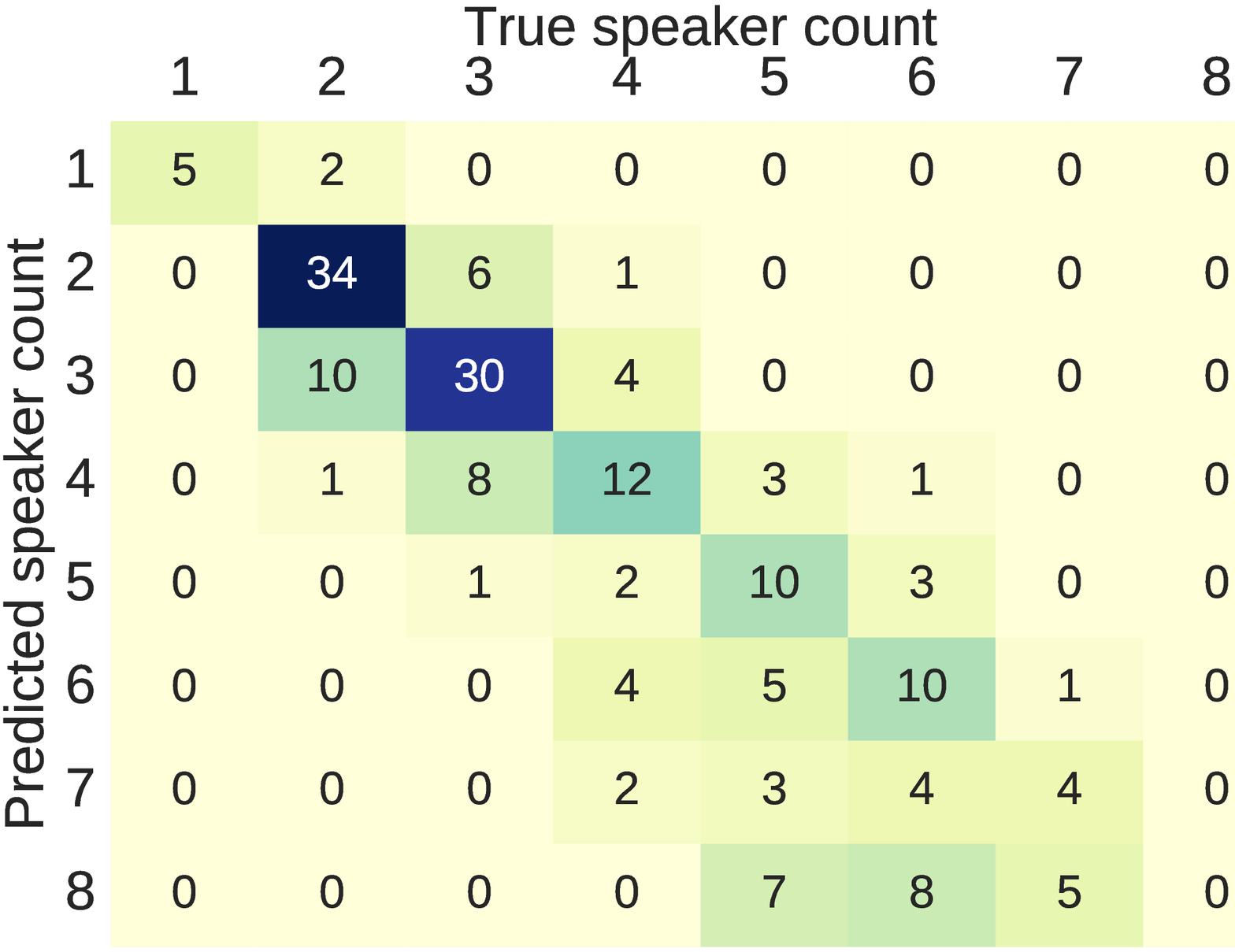}
\caption{Confusion matrix with variable number of speakers.}
\label{fig:conf}
\end{figure}

To test the diarization model's performance for a variable number of speakers, we trained the TDCN-SA+local+speaker model on 100K, 90 second meetings with $1$ to $8$ speakers. The 1-8 speaker test set uses LibriSpeech test data, and the observed DER on this test set is 19\%. Figure \ref{fig:conf} shows a confusion matrix for estimated speaker count from the diarization prediction. Note that, for lower numbers of speakers ($< 5$), the model predicts speaker count more accurately.

\vspace{-.5cm}
\section{Conclusion}
\label{sec:concl}
We introduced an end-to-end diarization model that performs meeting diarization in a single inference step and improves upon previous proposals by incorporating a local loss and a speaker detection module, a local-global network, and using a sequential iteration. The model is shown to work well on simulated meeting data as well as the LibriCSS dataset which contains real acoustic mixing. Future work involves improving the robustness and generalization of the model and testing on more realistic meeting data.

{\bf Acknowledgements:} Thanks to Zhuo Chen and Yi Luo for preparing the tools to generate LibriMeet-100K format training and test data and Quan Wang for valuable discussions.
A part of this work was started at JSALT 2020 at JHU, with support from Microsoft, Amazon and Google.

\vfill\pagebreak

\bibliographystyle{IEEEbib}
\bibliography{ref}

\begin{thebibliography}{10}

\bibitem{tranter2006overview}
Sue~E Tranter and Douglas~A Reynolds,
\newblock ``An overview of automatic speaker diarization systems,''
\newblock {\em {IEEE/ACM} TASLP}, vol. 14, no. 5, pp. 1557--1565, 2006.

\bibitem{anguera2012speaker}
Xavier Anguera, Simon Bozonnet, Nicholas Evans, Corinne Fredouille, Gerald
  Friedland, and Oriol Vinyals,
\newblock ``Speaker diarization: A review of recent research,''
\newblock {\em {IEEE/ACM} TASLP}, vol. 20, no. 2, pp. 356--370, 2012.

\bibitem{ryant2019second}
Neville Ryant, Kenneth Church, Christopher Cieri, Alejandrina Cristia, Jun Du,
  Sriram Ganapathy, and Mark Liberman,
\newblock ``The second {DIHARD} diarization challenge: Dataset, task, and
  baselines,''
\newblock {\em Proc. Interspeech}, pp. 978--982, 2019.

\bibitem{el2019joint}
Laurent El~Shafey, Hagen Soltau, and Izhak Shafran,
\newblock ``Joint speech recognition and speaker diarization via sequence
  transduction,''
\newblock {\em Proc. Interspeech}, pp. 396--400, 2019.

\bibitem{Chen2020libricss}
Z.~{Chen}, T.~{Yoshioka}, L.~{Lu}, T.~{Zhou}, Z.~{Meng}, Y.~{Luo}, J.~{Wu},
  X.~{Xiao}, and J.~{Li},
\newblock ``Continuous speech separation: Dataset and analysis,''
\newblock in {\em Proc. {ICASSP}}, 2020, pp. 7284--7288.

\bibitem{medennikov2020target}
Ivan Medennikov, Maxim Korenevsky, Tatiana Prisyach, Yuri Khokhlov, Mariya
  Korenevskaya, Ivan Sorokin, Tatiana Timofeeva, Anton Mitrofanov, Andrei
  Andrusenko, Ivan Podluzhny, et~al.,
\newblock ``Target-speaker voice activity detection: a novel approach for
  multi-speaker diarization in a dinner party scenario,''
\newblock {\em arXiv preprint arXiv:2005.07272}, 2020.

\bibitem{ccetin2006overlap}
{\"O}zg{\"u}r {\c{C}}etin and Elizabeth Shriberg,
\newblock ``Overlap in meetings: Asr effects and analysis by dialog factors,
  speakers, and collection site,''
\newblock in {\em Proc. Int'l Workshop on Machine Learning for Multimodal
  Interaction}. Springer, 2006, pp. 212--224.

\bibitem{shum2013unsupervised}
Stephen~H Shum, Najim Dehak, R{\'e}da Dehak, and James~R Glass,
\newblock ``Unsupervised methods for speaker diarization: An integrated and
  iterative approach,''
\newblock {\em {IEEE/ACM} TASLP}, vol. 21, no. 10, pp. 2015--2028, 2013.

\bibitem{sell2014speaker}
Gregory Sell and Daniel Garcia-Romero,
\newblock ``Speaker diarization with plda i-vector scoring and unsupervised
  calibration,''
\newblock in {\em Proc. SLT}. IEEE, 2014, pp. 413--417.

\bibitem{senoussaoui2013study}
Mohammed Senoussaoui, Patrick Kenny, Themos Stafylakis, and Pierre Dumouchel,
\newblock ``A study of the cosine distance-based mean shift for telephone
  speech diarization,''
\newblock {\em {IEEE/ACM} TASLP}, vol. 22, no. 1, pp. 217--227, 2013.

\bibitem{dehak2010front}
Najim Dehak, Patrick~J Kenny, R{\'e}da Dehak, Pierre Dumouchel, and Pierre
  Ouellet,
\newblock ``Front-end factor analysis for speaker verification,''
\newblock {\em {IEEE/ACM} TASLP}, vol. 19, no. 4, pp. 788--798, 2010.

\bibitem{wang2018speaker}
Quan Wang, Carlton Downey, Li~Wan, Philip~Andrew Mansfield, and Ignacio~Lopz
  Moreno,
\newblock ``Speaker diarization with {LSTM},''
\newblock in {\em Proc. {ICASSP}}. IEEE, 2018, pp. 5239--5243.

\bibitem{snyder2019xvector}
D.~{Snyder}, D.~{Garcia-Romero}, G.~{Sell}, A.~{McCree}, D.~{Povey}, and
  S.~{Khudanpur},
\newblock ``Speaker recognition for multi-speaker conversations using
  x-vectors,''
\newblock in {\em Proc. {ICASSP}}, 2019, pp. 5796--5800.

\bibitem{fujita2019}
Yusuke Fujita, Naoyuki Kanda, Shota Horiguchi, Kenji Nagamatsu, and Shinji
  Watanabe,
\newblock ``{End-to-End Neural Speaker Diarization with Permutation-Free
  Objectives},''
\newblock in {\em Proc. Interspeech}, 2019, pp. 4300--4304.

\bibitem{fujita2019end}
Yusuke Fujita, Naoyuki Kanda, Shota Horiguchi, Yawen Xue, Kenji Nagamatsu, and
  Shinji Watanabe,
\newblock ``End-to-end neural speaker diarization with self-attention,''
\newblock in {\em Proc. ASRU}. IEEE, 2019, pp. 296--303.

\bibitem{garcia2017speaker}
Daniel Garcia-Romero, David Snyder, Gregory Sell, Daniel Povey, and Alan
  McCree,
\newblock ``Speaker diarization using deep neural network embeddings,''
\newblock in {\em Proc. {ICASSP}}. IEEE, 2017, pp. 4930--4934.

\bibitem{wan2018generalized}
Li~Wan, Quan Wang, Alan Papir, and Ignacio~Lopez Moreno,
\newblock ``Generalized end-to-end loss for speaker verification,''
\newblock in {\em Proc. {ICASSP}}. IEEE, 2018, pp. 4879--4883.

\bibitem{huang2020speaker}
Zili Huang, Shinji Watanabe, Yusuke Fujita, Paola Garc{\'\i}a, Yiwen Shao,
  Daniel Povey, and Sanjeev Khudanpur,
\newblock ``Speaker diarization with region proposal network,''
\newblock in {\em Proc. {ICASSP}}. IEEE, 2020, pp. 6514--6518.

\bibitem{Ding2020}
Shaojin Ding, Quan Wang, Shuo-Yiin Chang, Li~Wan, and Ignacio~Lopez Moreno,
\newblock ``{Personal VAD: Speaker-Conditioned Voice Activity Detection},''
\newblock in {\em Proc. Odyssey 2020 Speaker and Language Recognition
  Workshop}, 2020, pp. 433--439.

\bibitem{von2019all}
Thilo von Neumann, Keisuke Kinoshita, Marc Delcroix, Shoko Araki, Tomohiro
  Nakatani, and Reinhold Haeb-Umbach,
\newblock ``All-neural online source separation, counting, and diarization for
  meeting analysis,''
\newblock in {\em Proc. {ICASSP}}. IEEE, 2019, pp. 91--95.

\bibitem{fujita2020neural}
Yusuke Fujita, Shinji Watanabe, Shota Horiguchi, Yawen Xue, Jing Shi, and Kenji
  Nagamatsu,
\newblock ``Neural speaker diarization with speaker-wise chain rule,''
\newblock {\em arXiv preprint arXiv:2006.01796}, 2020.

\bibitem{horiguchi2020end}
Shota Horiguchi, Yusuke Fujita, Shinji Watanabe, Yawen Xue, and Kenji
  Nagamatsu,
\newblock ``End-to-end speaker diarization for an unknown number of speakers
  with encoder-decoder based attractors,''
\newblock {\em arXiv preprint arXiv:2005.09921}, 2020.

\bibitem{e2ediarizationpatent}
Quan Wang, Yash Sheth, Ignacio~Lopez Moreno, and Li~Wan,
\newblock ``Speaker diarization using an end-to-end model,'' Google Patents,
  2019.

\bibitem{fujita2020end}
Yusuke Fujita, Shinji Watanabe, Shota Horiguchi, Yawen Xue, and Kenji
  Nagamatsu,
\newblock ``End-to-end neural diarization: Reformulating speaker diarization as
  simple multi-label classification,''
\newblock {\em arXiv preprint arXiv:2003.02966}, 2020.

\bibitem{zhang2019fully}
Aonan Zhang, Quan Wang, Zhenyao Zhu, John Paisley, and Chong Wang,
\newblock ``Fully supervised speaker diarization,''
\newblock in {\em Proc. {ICASSP}}. IEEE, 2019, pp. 6301--6305.

\bibitem{li2019discriminative}
Qiujia Li, Florian~L Kreyssig, Chao Zhang, and Philip~C Woodland,
\newblock ``Discriminative neural clustering for speaker diarisation,''
\newblock {\em arXiv preprint arXiv:1910.09703}, 2019.

\bibitem{Luo2019tasnet}
Y.~{Luo} and N.~{Mesgarani},
\newblock ``Conv-tasnet: Surpassing ideal time–frequency magnitude masking
  for speech separation,''
\newblock {\em {IEEE/ACM} TASLP}, vol. 27, no. 8, pp. 1256--1266, 2019.

\bibitem{vaswani2017attention}
Ashish Vaswani, Noam Shazeer, Niki Parmar, Jakob Uszkoreit, Llion Jones,
  Aidan~N Gomez, {\L}ukasz Kaiser, and Illia Polosukhin,
\newblock ``Attention is all you need,''
\newblock in {\em Advances in Neural Information Processing Systems}, 2017, pp.
  5998--6008.

\bibitem{katharopoulos2020transformers}
Angelos Katharopoulos, Apoorv Vyas, Nikolaos Pappas, and Fran{\c{c}}ois
  Fleuret,
\newblock ``Transformers are {RNN}s: Fast autoregressive transformers with
  linear attention,''
\newblock {\em arXiv preprint arXiv:2006.16236}, 2020.

\bibitem{clevert2016elu}
Djork-Arn{\'e} Clevert, Thomas Unterthiner, and Sepp Hochreiter,
\newblock ``Fast and accurate deep network learning by exponential linear units
  (elus),''
\newblock {\em arXiv preprint arXiv:1511.07289}, vol. 2, 2016.

\bibitem{bredin2019pyannoteaudio}
Hervé Bredin, Ruiqing Yin, Juan~Manuel Coria, Gregory Gelly, Pavel Korshunov,
  Marvin Lavechin, Diego Fustes, Hadrien Titeux, Wassim Bouaziz, and
  Marie-Philippe Gill,
\newblock ``pyannote.audio: neural building blocks for speaker diarization,''
  2019.

\end{thebibliography}

\end{document}